# Effect of Numerically Controlled Oscillator Bit Width in Phase Meters

Yuan-Ze Jiang[†], Yu-Jie Feng, Liu-Yang Chen, Bai-Fu Lu, Qi Xia, Ze-Bing Zhou and Yu-Rong Liang[*]

[1]MOE Key Laboratory of Fundamental Physical Quantities Measurement and Hubei Key Laboratory of Gravitation and Quantum Physics, PGMF, School of Physics, Huazhong University of Science and Technology, Wuhan 430074, China

**ABSTRACT**. Projects aiming to detect gravitational waves (GWs) in space in the millihertz range will utilize interferometers to measure the separations between free-falling test masses. The phasemeter measures the phase changes of the interference signals caused by the test masses' relative movements. The measurement sensitivity of the phasemeter is one of the key factors in the detection. In this work, we reviewed the core metrology of the phasemeter and evaluated the ultra-low noise performance of the phasemeter with analog signals. Frequency readout noise related to the bit width of the numerically controlled oscillator (NCO) inside the phasemeter is identified as one of the main noise sources of phase measurement theoretically and experimentally. After increasing the NCO bit widths, the single-channel phase noise of the phasemeter reached 2.0 μrad/Hz$^{1/2}$ at 6 mHz, and the differential phase noise reached 0.4 μrad/Hz$^{1/2}$ at 6 mHz. The phase noise performances remained consistent within the carrier frequency range of 4.9 MHz to 25.1 MHz.

## I. INTRODUCTION

The missions for space-borne GW detection include LISA [1–4], TianQin [5,6], Taiji [7,8], etc., which aim to observe GWs around the mHz band. They will use laser interferometry to accurately measure the relative displacement changes between the test masses on different satellites. Their requirements for the interferometric displacement noises are 15 pm/Hz$^{1/2}$ [4], 1 pm/Hz$^{1/2}$ [5], and 8 pm/Hz$^{1/2}$ [8], respectively.

The phasemeter reads the phase changes of the laser interferometric signals. Hence, its low-frequency phase measurement sensitivity is one of the key factors in GW detection. According to noise allocations, the phase measurement sensitivity of the phasemeter needs to be in the order of μrad/Hz$^{1/2}$ around the mHz band.

Early phase measurement solutions for space-borne GW detection include the zero-crossing method [9] and the IQ demodulation method [8]. After 2006, JPL [10] and AEI [11] began to develop phasemeters for LISA based on the field programmable gate array (FPGA) platform. Their works described the principle and basic structure of a digital phasemeter based on a digital phase-locked loop (DPLL), and the preliminary results of pure digital tests and external signal source tests were given.

After decades of development, the performance of the phasemeter has made great progress [12–16], but the research on the measurement sensitivity of the phasemeter is still an important work. The whole space-borne GW detector is very complicated, and the low-frequency noises coming from various instruments eventually turn into phase noises. To distinguish between different noise sources, it is first necessary to ensure the precision of the phasemeter is qualified.

The test methods of the phasemeter can be divided into single-signal phase measurement, two-signal differential phase measurement, and three-signal phase measurement [17,18]. To deduct the phase noise introduced by the signal source, most of the tests for the phasemeter give the result of differential measurement. However, the differential measurement cannot distinguish the common mode noises between the phasemeter channels. In this paper, we use a signal generator to produce RF signals, test the phasemeter by single-signal phase measurement, focus on exploring the low-frequency phase noise performance of the phasemeter, and present the single-channel and differential phase noises of the phasemeter.

We first review the basic principle of the phasemeter used for space-borne GW detection and analyze the main phase noises of the phasemeter. According to the analysis results, we find that the frequency readout noise of the DPLL is one of the main phase measurement noises. We used analog signals produced by a waveform generator to test the phasemeter with different bit widths of the NCOs in the DPLL and verified the influence of the frequency readout noise on the phase measurement. After increasing the NCO bit widths and proper temperature control of the experiment instruments, the single-channel phase noise of the phasemeter reached 2.0 μrad/Hz$^{1/2}$ at 6 mHz, and the differential phase noise reached 0.4

[†]Contact author: jiangyuanze@hust.edu.cn
*Contact author: liangyurong20@hust.edu.cn

μrad/Hz$^{1/2}$ at 6 mHz. The noise performance remained consistent within the carrier frequency range of 4.9 MHz to 25.1 MHz.

## II. PHASE MEASUREMENT PRINCIPLE AND NOISES

The basic phase readout unit for inter-satellite laser interferometry is shown in FIG 1. In inter-satellite laser interferometry, the main functions of the phasemeter include tracking the frequency and phase of the interferometric signals, controlling the beam direction according to differencial wavefront sensing [19,20], forming an optical phase-locked loop on slave satellites [21,22], and outputting the phase data in real-time on master satellites [18,23]. The FPGA platform is used as the hardware core to build a phasemeter, which can acquire and output phase information of multiple RF signals synchronously.

During the phase measurement process, the interferometric signal produced by the photoreceiver needs to be digitized by an analog-to-digital converter (ADC) with sufficient precision and rate. The sampling clock jitter noise during the analog-to-digital (A/D) conversion process disturbs the measurement and is eliminated by pilot tone correction [24]. The output of the photoreceiver is first superimposed with a pilot tone signal, and then the two are converted from analog signals to digital signals through an ADC. The digital signals are phase-tracked in real-time by the DPLLs inside the FPGA.

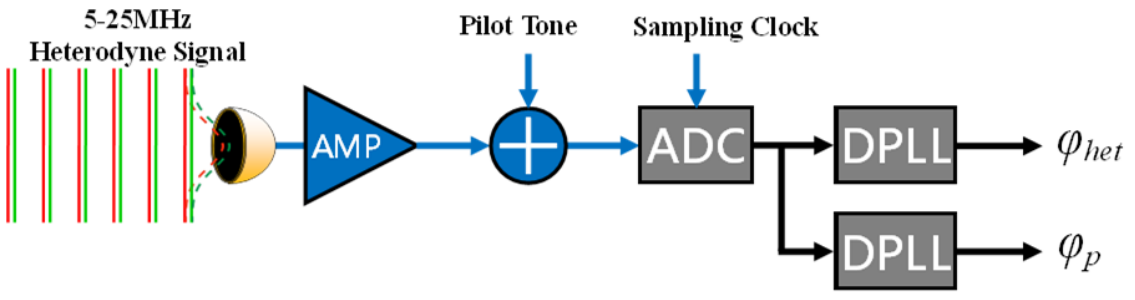

FIG 1. Schematic of interferometric phase measurement. A photoreceiver transfers the heterodyne optical signal to the RF signal. An ADC digitizes the combination of the heterodyne RF signal and the pilot tone signal. After proper filtering, the phases of the two signals are demodulated by two independent DPLLs.

FIG 2 shows the internal structure of a DPLL. Inside a DPLL, the input signal mixes with a sinusoidal signal of the same frequency (or frequency close to the signal frequency) produced by an NCO using a look-up table (LUT). The near-DC signal obtained by mixing is fed back to the NCO, which uses a phase increment register (PIR) and a phase accumulator (PA) to correct its frequency. Through this process, real-time frequency and phase tracking of the input signal can be realized. The sampling clock jitter noise of ADC is eliminated through the pilot tone correction process by using phase signals $\varphi_{het}$ and $\varphi_p$ produced by two sets of DPLLs. After this, the phase information of the interference signal is obtained.

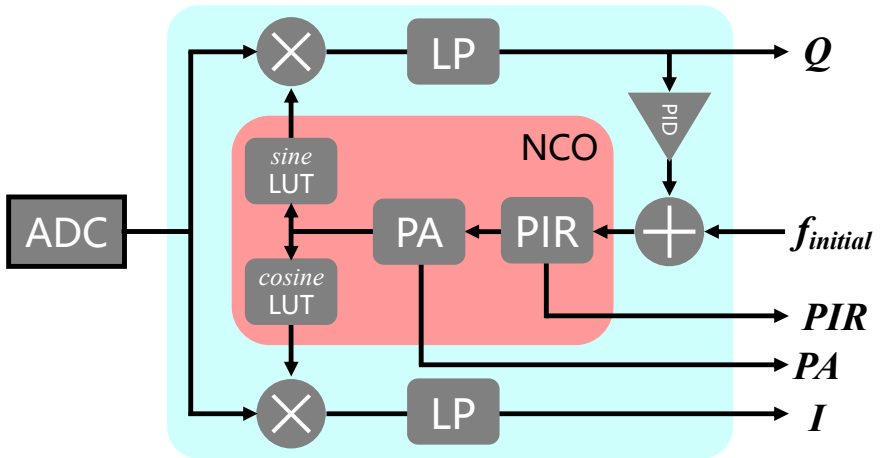

FIG 2. Diagram of the DPLL inside a phasemeter. The digitized signal mixes with a sinusoidal signal produced by an NCO. The near-DC signal obtained by mixing is fed back to the NCO, which uses a PIR and a PA to correct its output frequency. The phase can be reconstructed by the output of PIR.

The phasemeter's main measurement noises include A/D conversion quantization noise, sampling clock jitter noise, and frequency readout noise of the DPLL. The theoretical models of the above three noises are established below, and the noise contribution estimations around the mHz frequency band are given afterward.

### A. A/D conversion quantization noise

In the process of A/D conversion, limited resolution will introduce ADC quantization noise [25], and the relationship between this noise and relevant parameters is shown as follows:

$$\varphi_{ADC} = \frac{2^{-Q}}{\sqrt{6\cdot f_s}}\cdot\frac{V_{FS}}{V_{in}}\ \text{[rad/Hz}^{1/2}\text{]} \tag{1}$$

Where $Q$ represents the effective number of bits (ENOB) of ADC, $f_s$ represents the sampling clock frequency, $V_{FS}$ represents the voltage range of ADC, and $V_{in}$ represents the voltage amplitude of the input signal. This noise shows up as white noise in phase measurement. Assuming that the ENOB is 12 bits, the sampling clock is 80 MHz, and the input signal is full-scale, the A/D conversion quantization noise value is about $1.1\times10^{-8}$rad/Hz$^{1/2}$.

### B. Sampling clock jitter noise

In the process of digitization, the fluctuation of the sampling clock will introduce additional phase noise. The fluctuation comes from the clock source jitter, the disturbance caused by the clock transfer lines, and the ADC timing jitter. Although the clock source jitters will be removed by the TDI process in space-borne GW detectors, it disturbs the evaluation of the phasemeter performance. Taking the fluctuation of a

†Contact author: jiangyuanze@hust.edu.cn
*Contact author: liangyurong20@hust.edu.cn

typical constant temperature crystal oscillator [26] as an example:

$$\tau(f) = (3.6\times10^{-33} + \frac{1.6\times10^{-27}}{f} + \frac{10^{-26}}{f^4})^{1/2}\ [\mathrm{s/Hz^{1/2}}] \quad (2)$$

If we use this crystal oscillator as the sampling clock, the phase of the signal to be measured will superimpose phase noise:

$$\varphi_{het} = \varphi_0 + 2\pi f_{het} \int \tau(t)dt\ [\mathrm{rad}] \quad (3)$$

$\varphi_0$ represents the interferometric phase signal to be measured, $\varphi_{het}$ represents the phase signal after $\varphi_0$ superimposed with the sampling clock jitter noise, $f_{het}$ represents the frequency of the interferometric signal, and $\tau(t)$ represents the sampling clock jitter in the time domain.

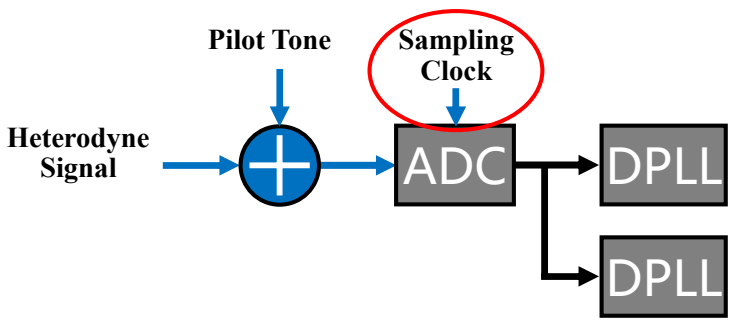

FIG 3. Pilot tone correction to eliminate sampling clock jitter noise. The sampling clock jitter affects the heterodyne signal and the pilot tone signal simultaneously, which can be eliminated by pilot tone correction using phases of the two signals.

Without pilot tone correction, phase noise of about 30 rad/Hz$^{1/2}$ at 6 mHz will be added to the 10 MHz single channel signal by the sampling clock, which will completely disturb the measurement. If a pilot signal is added to the interferometric signal before the digitization, both the interferometric signal and the pilot signal will be affected by the sampling clock jitter, as shown in FIG 3. The phase of the pilot signal can be used to eliminate the sampling clock jitter noise.

$$\varphi_p = 2\pi f_p \int \tau(t)dt\ [\mathrm{rad}] \quad (4)$$

$$\varphi_{het}^{cor} = \varphi_{het} - \frac{f_{het}}{f_p}\varphi_p\ [\mathrm{rad}] \quad (5)$$

$\varphi_p$ represents the sampling clock jitter noise superimposed in the pilot signal, $f_p$ represents the frequency of the pilot signal, and $\varphi_{het}^{cor}$ represents interferometric phase signal after pilot correction.

### C. Frequency readout noise of the DPLL

Inside the NCO, the frequency word produced by the PIR is accumulated and truncated by the PA and then entered into the LUT to keep the loop locked. Meanwhile this frequency word is down sampled and outputted in real-time and the phase can be obtained by integrating the frequency.

The truncation noise of the frequency is one of the noises inside the phase-locked loop and it will be suppressed by the error function of the DPLL [12,27]. In the process of frequency output, similar to ADC quantization noise, the limited bit width of the frequency signal will introduce frequency readout quantization noise. Previous works [25,28] have introduced this noise source. We take the down-sampling process of signal output into account, and the relationship between this noise and relevant parameters is shown as follows:

$$\varphi_{PIR} = \frac{1}{f_{fourier}} \cdot \frac{f_s/2^M}{\sqrt{6\cdot f_s/2^W}}\ [\mathrm{rad/Hz^{1/2}}] \quad (6)$$

$f_{fourier}$ is the Fourier frequency, $M$ is the bit width of the frequency word in the DPLL, $f_s$ is the sampling clock, and $W$ is the down-sampling bit width. Frequency readout noise shows up in the phase noise as $1/f$ noise. Assuming a digital frequency bit width of 48 bits, a down-sampling coefficient of $2^{22}$, and a sampling clock of 80 MHz, the contribution of this noise is 4.5 μrad/Hz$^{1/2}$ at 6 mHz.

Note that both the interferometric phase and the pilot phase are affected by this noise. However, the two noise contributions are uncorrelated and they should be summed quadratically with the pilot tone correction coefficient $f_{het}/f_p$ taken into consideration as shown in equation (7), hence the frequency readout noise of the interferometric phase is more significant after pilot tone correction.

$$\varphi_{PIR,sum} = \sqrt{(\varphi_{PIR,het})^2 + (\frac{f_{het}}{f_p}\varphi_{PIR,p})^2}\ [\mathrm{rad/Hz^{1/2}}] \quad (7)$$

Usually, $f_p$ is larger than $f_{het}$. Assuming $f_{het}$ = 10 MHz and $f_p$ = 30 MHz, the frequency readout noise will increase by about 5% in the corrected signal because of the influence of the pilot tone.

### D. Noise summary

The noises mentioned above in a single-channel phase signal before pilot tone correction are summarized in FIG 4.

The red dot-dashed line represents the sampling clock jitter noise, the green dotted line represents the frequency readout noise of NCO, and the blue solid line represents the quantization noise of ADC. The black dot-dashed line represents the phase noise corresponding to the single sideband (SSB) phase noise of the input signal when the SSB phase noise is -140 dBc/Hz. The pink dot-dashed line represents the temperature-phase coupled noise estimated by the literature [29], and the orange dashed line is a typical phase measurement noise requirement of 0.2 pm/Hz$^{1/2}$ with noise shape function NSF = $[1+(6\ \mathrm{mHz}/f)^4]^{1/2}$.

†Contact author: jiangyuanze@hust.edu.cn
*Contact author: liangyurong20@hust.edu.cn

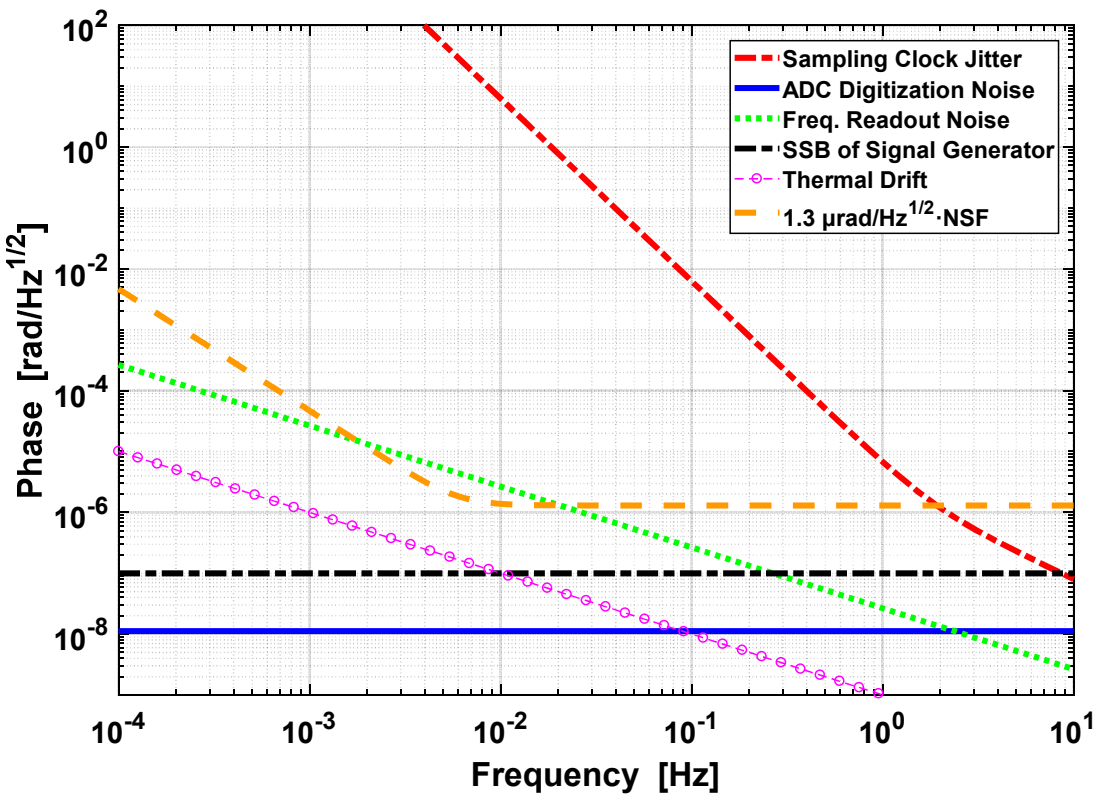

FIG 4. Theoretical noises of a single-channel phase signal before pilot tone correction. The sampling clock jitter will completely disturb the measurement without pilot tone correction. The frequency readout noise of the DPLL will also limit the performance if the bit width of the NCO is insufficient.

According to the theoretical prediction, after the pilot tone correction, the major noise of the system will be the frequency readout noise of the DPLL. We present the experimental verification results of this noise in the following sections.

## III. EXPERIMENTAL DESIGN

In space-borne GW detection, each satellite carries an ultra-stable oscillator (USO) as a clock source for pilot signals, heterodyne phase-locking, clock sideband modulation, etc. The interferometric and pilot tone signals may contain clock jitter noises of different USOs. These clock jitter noises will be deducted during the TDI process [30,31] by the clock synchronization technique [32,33].

To evaluate the performance of the phasemeter itself, the issue of clock synchronization is ignored in this experiment scheme, and the same USO is used to generate the interferometric signal and the pilot tone signal to evaluate the noise floor of the phasemeter under electrical conditions.

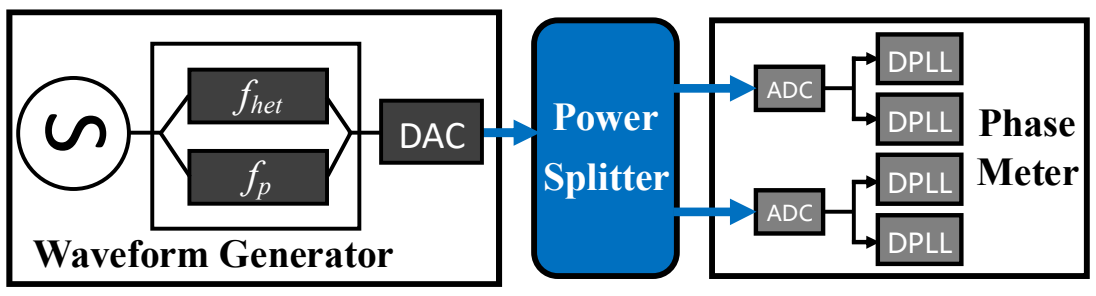

FIG 5. Diagram of the phasemeter evaluation setup. The waveform generator produces two signals with different frequencies and outputs them to one BNC plug. The multi-frequency signal is then split and digitized by two ADC. Finally, the phasemeter obtains four phases from the input signals.

The experimental scheme is shown in FIG 5. To ensure the homology of the interferometric signal and the pilot signal as much as possible, we choose the waveform generator 33622A of Keysight to generate signals after testing various signal sources. Under its 'combined' mode, two signals are added in the internal digital program and transferred to the same digital-to-analog converter (DAC) module. Channel 1 generates radio frequency signals of 4.9 to 25.1 MHz to represent the interferometric signals, and channel 2 generates a radio frequency signal of 34.1 MHz as the pilot signal.

The multi-frequency signal is divided into two channels by the power divider. The divider's outputs are connected to the two ADC measurement channels of the phasemeter. After digitization, appropriate filtering, phase-locking, and decimation, the two interferometric frequencies and the two pilot frequencies are transferred from the phasemeter to the LabView program utilized on a computer by a serial port. The LabView program implements frequency integration, pilot correction, and data recording.

To eliminate the influence of environmental noises, we apply appropriate temperature control on the experimental equipment, including active or passive temperature control on the waveform generator and phasemeter, and reduce the influence of temperature gradient between the channels of the phasemeter by using heat-conducting fins. To avoid the disturbances introduced by RF cables, we use the shortest possible RF cables and ensure that the cable lengths are the same for each phasemeter channel.

## IV. EXPERIMENTAL RESULTS

### A. Pilot tone correction

To evaluate the pilot tone correction performance of the phasemeter, the interferometric signal was set to 10.1 MHz, the pilot signal was set to 34.1 MHz, the sampling clock of the phasemeter was set to 80 MHz, the down-sampling coefficient was set to $2^{22}$, and the bit widths of the NCOs was set to 54 bits. The single-channel phase measurement performance of the phasemeter is shown in FIG 6.

From the top down, the solid lines represent the phases of the two 34.1 MHz pilot tone signals, the dot-dashed lines represent the phases of the two 10.1 MHz interferometric signals, the orange dashed line represents the noise requirement, and the dashed lines at the bottom represent the pilot tone correction results of the two channels. As can be seen from the figure, the noise floor of the phasemeter reached 1.3 μrad/Hz$^{1/2}$*NSF after pilot tone correction, and the noise rejection ratio of pilot tone correction was greater than $10^5$ at 6 mHz. All the following

†Contact author: jiangyuanze@hust.edu.cn
*Contact author: liangyurong20@hust.edu.cn

experimental results are based on data after the pilot correction.

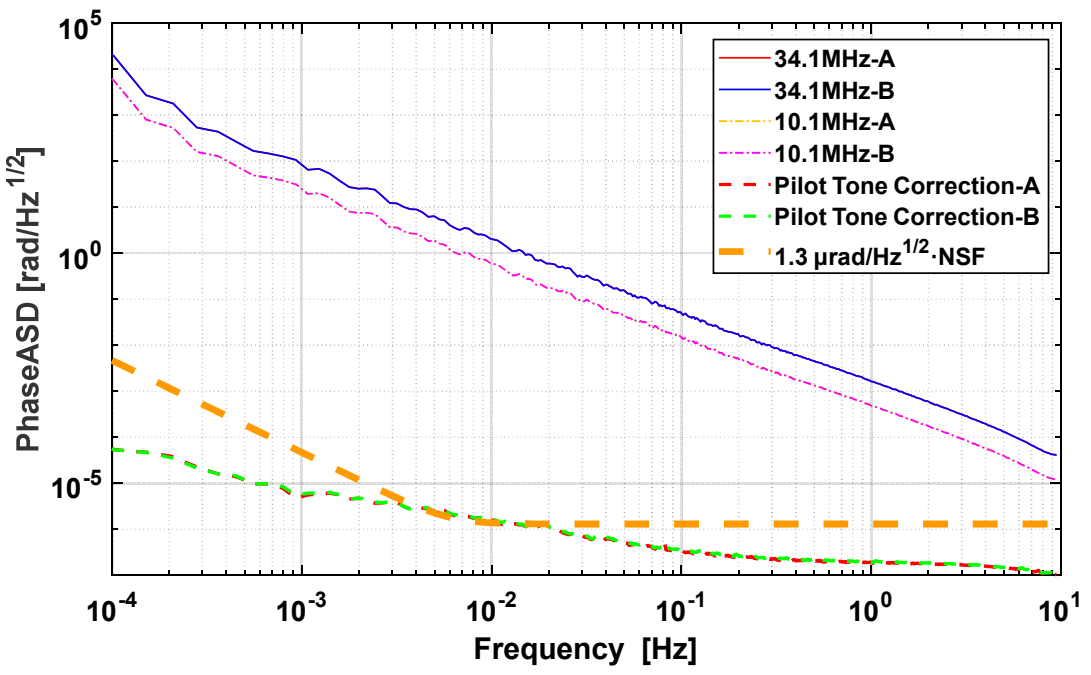

FIG 6. Pilot tone correction performance with interferometric frequency of 10.1 MHz and pilot tone frequency of 34.1 MHz. The noise floor of the phasemeter reached 1.3 μrad/Hz$^{1/2}$*NSF after pilot tone correction, and the noise rejection ratio of pilot tone correction was greater than $10^5$ at 6 mHz.

### B. Frequency readout noise modulation experiment

From the previous noise analysis, we find that the frequency readout noise of the DPLL is one of the main noises of the phasemeter, and the digital bit width of the NCO determines the frequency readout noise. We set the sampling clock of the phasemeter to 80 MHz and the down-sampling coefficient to $2^{22}$ and tested the phase measurement results of different NCO bit widths. The single-channel and differential results are shown in FIG 7 and FIG 8.

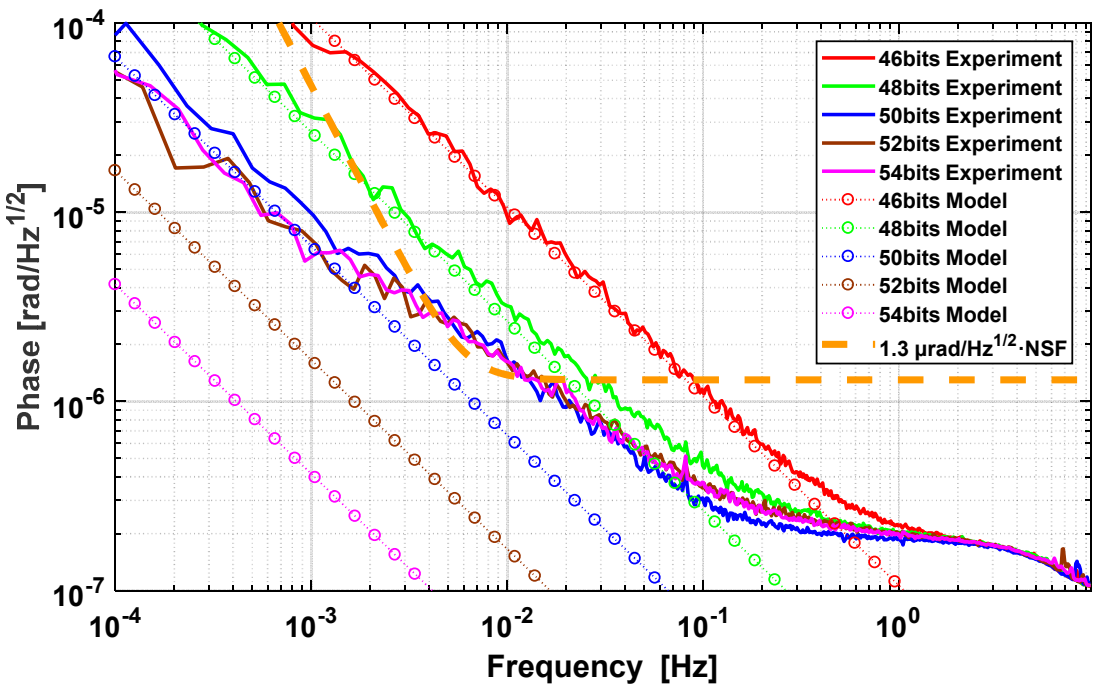

FIG 7. Single-channel phase noise floors with different NCO bit widths. The interferometric signal and the pilot tone signal remained the same with the pilot tone correction experiment mentioned above. The Verilog code of the phasemeter was recompiled for different NCO bit widths.

The solid lines are the experimental results under different NCO bit widths, the dotted lines marked with 'o' are the predicted values of the theoretical model, and the orange dashed line is the noise requirement.

†Contact author: jiangyuanze@hust.edu.cn
*Contact author: liangyurong20@hust.edu.cn

Results in FIG 7 indicated that the single-channel result was limited to 2.0 μrad/Hz$^{1/2}$ at 6 mHz.
When the NCO bit widths are greater than 50 bits, the noises caused by other factors begin to dominate at a frequency range lower than 1 Hz.

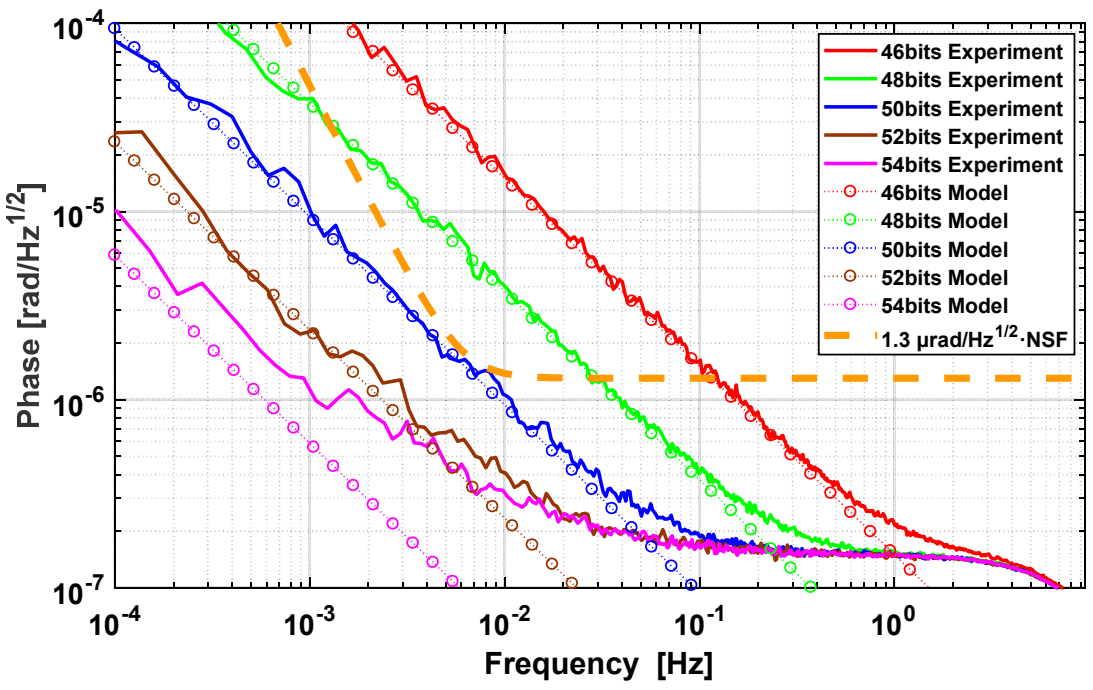

FIG 8. Differential phase noise floors with different NCO bit widths. The theoretical model predictions for different NCO bit widths in this figure were $\sqrt{2}$ times the corresponding values in FIG 7.

In FIG 8, the differential results of the two channels showed a certain amount of common mode suppression for the unknown noises. By further increasing the NCO bit widths to 54 bits, the final differential result reached 0.4 μrad/Hz$^{1/2}$ at 6mHz.

### C. Test results of different interferometric frequencies

FIG 9 and FIG 10 are phase measurements at different interferometric frequencies. The interferometric signal was set to 4.9-25.1 MHz, the pilot signal was set to 34.1 MHz, and the NCO bit widths were set to 54 bits to eliminate the frequency readout noise of the DPLL. Fig 9 shows the single-channel phase signals after pilot correction, and the phase measurement noise floors in different interferometric frequencies reached 2.0 μrad/Hz$^{1/2}$ at 6 mHz. Fig 10 shows the differential phase noise floors reached 0.4 μrad/Hz$^{1/2}$ at 6 mHz.

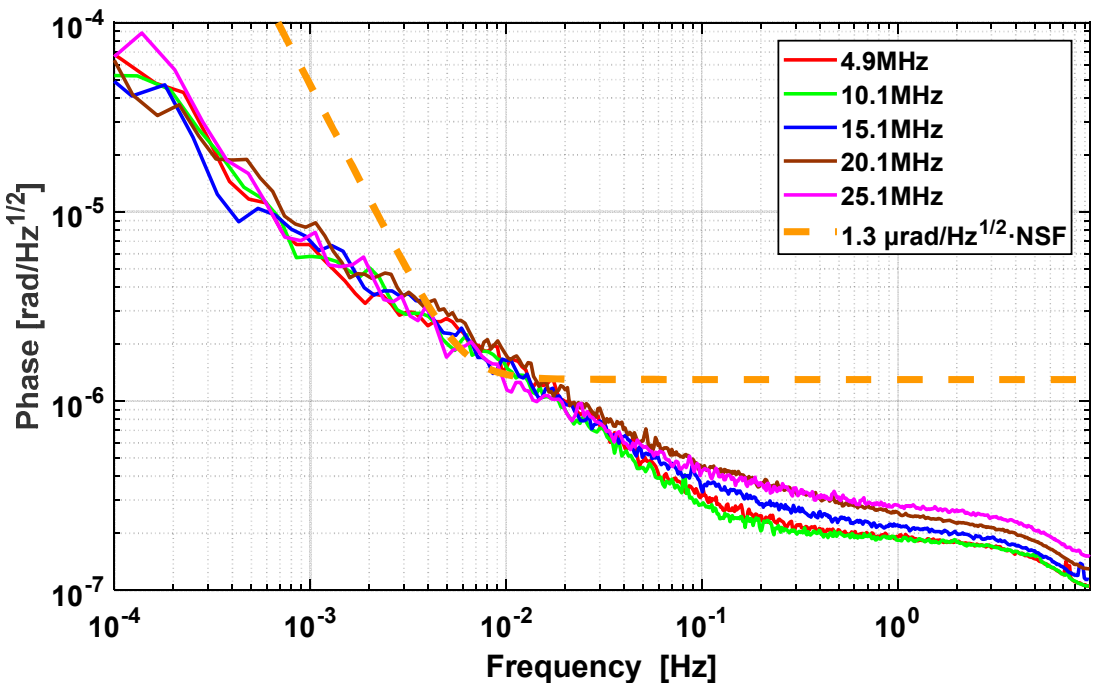

FIG 9. Single-channel phase noise floors of different carrier frequencies.

It can be seen from the results that the single-channel phase measurement results with the NCO bit widths of 54 bits have reached our requirement, and the noise floor performances of the phasemeter remained consistent under different interferometric frequencies.

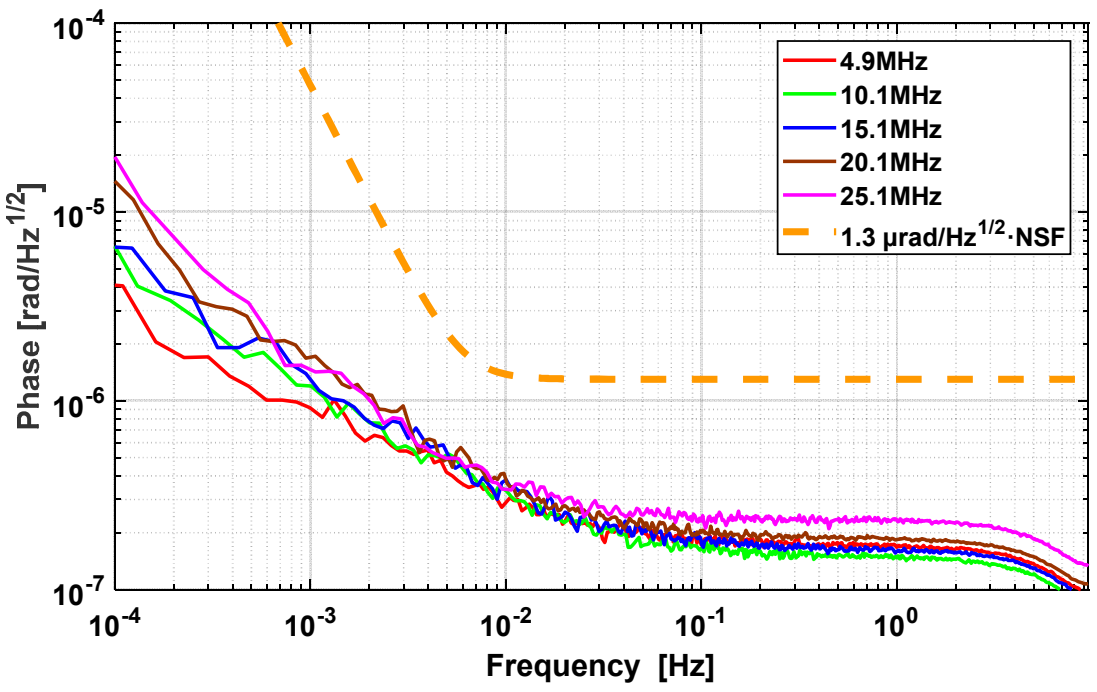

FIG 10. Differential phase noise floors of different carrier frequencies.

From the comparison of the single-channel and differential results, it can be seen that there are still some unknown common-mode noises between the two channels after pilot tone correction, which are independent of the interferometric frequency and NCO bit width. The noise sources may be USO stability, signal source DAC frequency nonlinearity, signal transmission line nonlinearity, and temperature response of the phasemeter's sensitive components.

## V. CONCLUSIONS

This paper reviewed the basic principle of the digital phasemeter used for GW detection and derived the main noises of the digital phasemeter under typical parameters. According to the calculation, frequency readout noise is one of the limiting factors of the noise floor performance of the phasemeter based on a DPLL. To verify the influence of this noise, we use the most compact structure to test the noise floor of the phasemeter.

We investigated the single-channel and differential phase noise floors of the phasemeter with different NCO bit widths using a waveform generator with suitable temperature control. The experiment results showed that under the condition of 54 bits NCO bit widths, the single-channel phase noise floors of the phasemeter reached 2.0 μrad/Hz$^{1/2}$ at 6 mHz, and the differential phase noise floors reached 0.4 μrad/Hz$^{1/2}$ at 6 mHz. The noise performances remained consistent in the carrier frequency range of 4.9 MHz to 25.1 MHz. Phase measurement performance is one of the baselines in space-borne GW detection, and the work in this paper can provide a reference for the design and evaluation of phase measurement, heterodyne phase locking, wavefront differential sensing, clock synchronization, and other functions in the electronic system of space-borne GW detection. We also expect this work to benefit fields requiring low phase noise such as fundamental physics [34,35], frequency stabilization [36,37], spacecraft attitude control [38], and navigation [39].

## ACKNOWLEDGMENTS

This work was supported by the National Key Research and Development Program of China (Grant Nos. 2022YFC2204001,2022YFC2203903).

†Contact author: jiangyuanze@hust.edu.cn
*Contact author: liangyurong20@hust.edu.cn

†Contact author: jiangyuanze@hust.edu.cn
*Contact author: liangyurong20@hust.edu.cn

[†]Contact author: jiangyuanze@hust.edu.cn
*Contact author: liangyurong20@hust.edu.cn